\documentclass[12pt]{article}
\usepackage{amsfonts}
\usepackage{latexsym,amsmath,amssymb}
\usepackage{graphics,epstopdf}
\usepackage[pdftex]{graphicx}
\numberwithin{equation}{section}

\newtheorem{thm}{Theorem}[section]

\numberwithin{equation}{section}

\begin{document}

\bigskip

\bigskip

\begin{center}
{\Large \textbf{A de Casteljau Algorithm for
 Bernstein type Polynomials based on $(p,q)$-integers
 }}

\bigskip

\textbf{Khalid Khan,}$^{1)}$ \textbf{D.K. Lobiyal}$^{1)}$ and  \textbf{Adem kilicman}$^{2)}$

$^{1)}$School of Computer and System Sciences, SC \& SS, J.N.U., New Delhi-110067., India%
\\[0pt]khalidga1517@gmail.com; dklobiyal@gmail.com\\
$^{2)}$Department of Mathematics,
Faculty of Science, University Putra
Malaysia, Malaysia \\[0pt]akilicman@putra.upm.edu.my \\[0pt]

\bigskip

\bigskip

\textbf{Abstract}
\end{center}

\parindent=8mm {\footnotesize {In this paper, a de Casteljau algorithm to compute $(p,q)$-Bernstein B$\acute{e}$zier curves based on $(p,q)$-integers is introduced. We study  the nature of degree elevation and degree reduction for $(p,q)$-B$\acute{e}$zier Bernstein functions. The new curves have some properties similar to $q$-B$\acute{e}$zier curves. Moreover, we construct the corresponding tensor product surfaces over the rectangular domain $(u, v) \in [0, 1] \times [0, 1] $ depending on four parameters. We also study the de Casteljau algorithm and degree evaluation properties of the surfaces for these generalization over the rectangular domain. Furthermore, some fundamental properties for $(p,q)$-Bernstein B$\acute{e}$zier curves are discussed. We get $q$-B$\acute{e}$zier curves and surfaces for $(u, v) \in [0, 1] \times [0, 1] $  when we set the parameter $p_1=p_2=1.$

\bigskip


{\footnotesize \emph{MSC: primary 65D17; secondary 41A10, 41A25, 41A36.}: \newline

\bigskip

\section{Introduction}

\parindent=8mm   Recently, Mursaleen et al \cite{mka1,mkar} applied $(p,q)$-calculus in
approximation theory and introduced the first $(p,q)$-analogue of Bernstein
operators based on $(p,q)$-integers. Motivated by the work of Mursaleen et al \cite{mka1,mkar}, the idea of $(p,q)$-calculus and its importance. We construct $(p,q)$-B$\acute{e}$zier curves and surfaces based on  $(p,q)$-integers which is further generalization of $q$-B$\acute{e}$zier curves and surfaces. For similar works based on $(p,q)$-integers, one can refer \cite{zmn,mur8,mka3,mnak1,mishra,wafi}.\\

\parindent=8mm It was  S.N. Bernstein \cite{brn} in 1912, who first introduced his famous operators $%
B_{n}: $ $C[0,1]\rightarrow C[0,1]$ defined for any $n\in \mathbb{N}$
and for any function $f\in C[0,1]$
\begin{equation}
B_{n}(f;x)=\sum\limits_{k=0}^{n}\left(
\begin{array}{c}
n \\
k%
\end{array}%
\right) x^{k}(1-x)^{n-k}f\biggl{(}\frac{k}{n}\biggl{)},~~x\in \lbrack 0,1].
\end{equation}
 and named it Bernstein polynomials to prove the Weierstrass
theorem \cite{pp}.
Later it was found that Bernstein polynomials possess many remarkable properties and has various applications in areas such as approximation theory \cite{pp}, numerical analysis,
computer-aided geometric design, and solutions of differential equations due to its fine properties of approximation \cite{hp}.

In  computer aided geometric design (CAGD), Bernstein polynomials and its variants are used in order to preserve the shape of the curves or surfaces. One of the most important curve in CAGD \cite{thomas} is the classical B$\acute{e}$zier curve \cite{Bezier} constructed with the help of Bernstein basis functions.

 In recent years, generalization of the B$\acute{e}$zier curve with shape parameters has received continuous attention.
 Several authors were concerned with the problem of changing the shape of curves and surfaces, while keeping the
control polygon unchanged and thus they generalized the B$\acute{e}$zier curves in \cite{farouki,wcq,hp}. \\

The rapid development of $q$-calculus \cite{vp} has led to the discovery of new generalizations of Bernstein polynomials
involving $q$-integers \cite{lp,ma1,hp} .
 The aim of these generalizations is to provide appropriate and powerful tools to application areas such as numerical analysis,
computer-aided geometric design, and solutions of differential equations.

\parindent=8mm In 1987, Lupa\c{s} \cite{lp} introduced the first $q$-analogue of Bernstein operator as follows

\begin{equation}\label{e9a}
L_{n,q}(f;x)= \sum\limits_{k=0}^{n}~~ \frac{f \bigg(\frac{[k]_{q}}{[n]_{q}}\bigg) ~\left[\begin{array}{c}
n \\
k%
\end{array}%
\right] _{q}  q^{\frac{k(k-1)}{2}}~x^{k}~(1-x)^{n-k}}{\prod\limits_{j=1}^{n}\{(1-x)+q^{j-1} x\}},
\end{equation}

and investigated its approximating and
shape-preserving properties.\\

 In 1996, Phillips \cite{pl} proposed another $q$-variant of the classical Bernstein operator, the so-called Phillips $q$-Bernstein operator  which attracted lots of investigations.

\begin{equation}
B_{n,q}(f;x)=\sum\limits_{k=0}^{n}\left[
\begin{array}{c}
n \\
k%
\end{array}%
\right] _{q}x^{k}\prod\limits_{s=0}^{n-k-1}(1-q^{s}x)~~f\left( \frac{%
[k]_{q}}{[n]_{q}}\right) ,~~x\in \lbrack 0,1]
\end{equation}
where $B_{n,q}: $ $C[0,1]\rightarrow C[0,1]$ defined for any $n\in \mathbb{N}$
and any function $f\in C[0,1].$ \\

The $q$-variants of Bernstein polynomials
provide one shape parameter for constructing free-form curves and surfaces, Phillips $q$-Bernstein operator was applied well
in this area.

 In 2003, Oruk and Phillips \cite{hp} used the basis functions of Phillips $q$-Bernstein operator for construction of
$q$-B$\acute{e}$zier curves, which they call Phillips $q$-B$\acute{e}$zier curves, and studied the properties of degree reduction and elevation.

Thus with the development of $(p,q)$-analogue of Bernstein operators and its variants, one natural question arises, how it can be used in order to preserve the shape of the curves or surfaces. In this way, it opens a new research direction which requires further investigations.\\

Before proceeding further, let us recall certain notations of $(p,q)$-calculus .\\

The $(p,q)$ integers $[n]_{p,q}$ are defined by
$$[n]_{p,q}=\frac{p^n-q^n}{p-q},~~~~~~~n=0,1,2,\cdots, ~~p>q>0.$$

The formula for $(p,q)$-binomial expansion is as follow:
\begin{equation*}
(ax+by)_{p,q}^{n}:=\sum\limits_{k=0}^{n}p^{\frac{(n-k)(n-k-1)}{2}}q^{\frac{k(k-1)}{2}}
\left[
\begin{array}{c}
n \\
k%
\end{array}%
\right] _{p,q}a^{n-k}b^{k}x^{n-k}y^{k},
\end{equation*}
$$(x+y)_{p,q}^{n}=(x+y)(px+qy)(p^2x+q^2y)\cdots (p^{n-1}x+q^{n-1}y),$$
$$(1-x)_{p,q}^{n}=(1-x)(p-qx)(p^2-q^2x)\cdots (p^{n-1}-q^{n-1}x),$$\\

where  $(p,q)$-binomial coefficients are defined by
$$\left[
\begin{array}{c}
n \\
k%
\end{array}%
\right] _{p,q}=\frac{[n]_{p,q}!}{[k]_{p,q}![n-k]_{p,q}!}.$$

Details on $(p,q)$-calculus can be found in \cite{jag,mka1,mah}.\\

 The $(p,q)$-Bernstein Operators introduced by Mursaleen et al is as follow:
\begin{equation}\label{e1}
B_{n,p,q}(f;x)=\frac{1}{p^{\frac{n(n-1)}{2}}}\sum\limits_{k=0}^{n}\left[
\begin{array}{c}
n \\
k%
\end{array}%
\right] _{p,q} p^{\frac{k(k-1)}{2}}x^{k}\prod\limits_{s=0}^{n-k-1}(p^s-q^{s}x)~~f\left( \frac{%
[k]_{p,q}}{p^{k-n}~~[n]_{p,q}}\right) ,~~x\in \lbrack 0,1]
\end{equation}\\

Note when $p=1,$ $(p,q)$-Bernstein Operators given by (\ref{e1}) turns out to be $q$-Bernstein Operators.

Also, we have
\begin{align*}
(1-x)^{n}_{p,q}&=\prod\limits_{s=0}^{n-1}(p^s-q^{s}x) =(1-x)(p-qx)(p^{2}-q^{2}x)...(p^{n-1}-q^{n-1}x)\\
&=\sum\limits_{k=0}^{n} {(-1)}^{k}p^{\frac{(n-k)(n-k-1)}{2}} q^{\frac{k(k-1)}{2}}\left[
\begin{array}{c}
n \\
k
\end{array}%
\right] _{p,q}x^{k}
\end{align*}

Again by some simple calculations and using the property of $(p,q)$-integers, we get $(p,q)$-analogue of Pascal's relation as follow:

\begin{equation}\label{e2}
\left[
\begin{array}{c}
n \\
k%
\end{array}%
\right] _{p,q}= q^{n-j}\left[
\begin{array}{c}
n-1 \\
k-1%
\end{array}%
\right] _{p,q}+ p^{j}\left[
\begin{array}{c}
n-1 \\
k%
\end{array}%
\right] _{p,q}
\end{equation}

\begin{equation}\label{e3}
\left[
\begin{array}{c}
n \\
k%
\end{array}%
\right] _{p,q}= p^{n-j}\left[
\begin{array}{c}
n-1 \\
k-1%
\end{array}%
\right] _{p,q}+ q^{j}\left[
\begin{array}{c}
n-1 \\
k%
\end{array}%
\right] _{p,q}
\end{equation}\\

We apply $(p,q)$-calculus and introduce first the $(p,q)$-B$\acute{e}$zier curves and surfaces based on $(p,q)$-integers which is further generalization of $q$-B$\acute{e}$zier curves and surfaces, for example, \cite{wcq,hp}.\\

The outline of this paper is as follow: Section $2$ introduces a $(p,q)$-analogue of the Bernstein functions $B^{k,n}_{p,q}$ and their Properties. Section $3$ introduces degree elevation and degree reduction properties for $(p,q)$-analogue of the Bernstein functions. Section $4$ introduces a de Casteljau
type algorithm  for $B^{k,n}_{p,q}$. In Section $5$ we define a tensor
product patch based on algorithm $1$ and its geometric properties as well as a degree
elevation technique are investigated. Furthermore tensor product of $(p,q)$-B$\acute{e}$zier surfaces on $[0, 1] \times [0, 1]$  for $(p,q)$-Bernstein polynomials are introduced and its properties that is inherited from the univariate case are being discussed.\\


\section{$(p,q)$-Bernstein functions}

The $(p,q)$-Bernstein functions is
as follows
\begin{equation}\label{ee4}
B^{k,n}_{p,q}(t)=\frac{1}{p^{\frac{n(n-1)}{2}}}\left[
\begin{array}{c}
n \\
k%
\end{array}%
\right] _{p,q} p^{\frac{k(k-1)}{2}}~~t^{k}(1-t)^{n-k}_{p,q} ,~~~~~t\in \lbrack 0,1]
\end{equation}\\

where
\begin{equation*}
(1-t)^{n-k}_{p,q}=\prod\limits_{s=0}^{n-k-1}(p^s-q^{s}t)
\end{equation*}

\subsection{Properties of the $(p,q)$-analogue of the Bernstein functions}

\begin{thm}
   The $(p,q)$-analogue of the Bernstein functions  possess the following properties:\\

 (1.) Non-negativity: $B^{k,n}_{p,q}(t)\geq 0$
 $k = 0, 1, . . . , n,~~~ t \in [0, 1].$ \\

(2.) Partition of unity:

\begin{equation*}
\sum\limits_{k=0}^{n} B^{k,n}_{p,q}(t)= 1, ~~~~\text{for every}~~ t \in [0, 1].
\end{equation*}\\

(3.)Both sided end-point property:
\begin{equation*}
B^{k,n}_{p,q}(0)=\left\{
\begin{array}{ll}
1 ,~~~~~\mbox{if $k=0$ } &  \\
&  \\
0,~~~~~~~~~~\mbox{$ k \neq 0$} &
\end{array}%
\right.
\end{equation*}

\begin{equation*}
B^{k,n}_{p,q}(1)=\left\{
\begin{array}{ll}
1,~~~~~~~~~\mbox{if $k=n$ } &  \\
&  \\
0,~~~~~~~~~~\mbox{$ k \neq n $} &
\end{array}%
\right.
\end{equation*}\\
when $p=1,$ then both side end point interpolation property holds.\\

%

(3.) Reducibility: when $p = 1,$ formula $(2.1)$ reduces to the $q$-Bernstein bases.\\
\end{thm}

\textbf{Proof:} All these property can be deduced easily from  equation (\ref{ee4}).\\

\newpage
\begin{figure*}[htb!]
\begin{center}
\includegraphics[height=6cm, width=8cm]{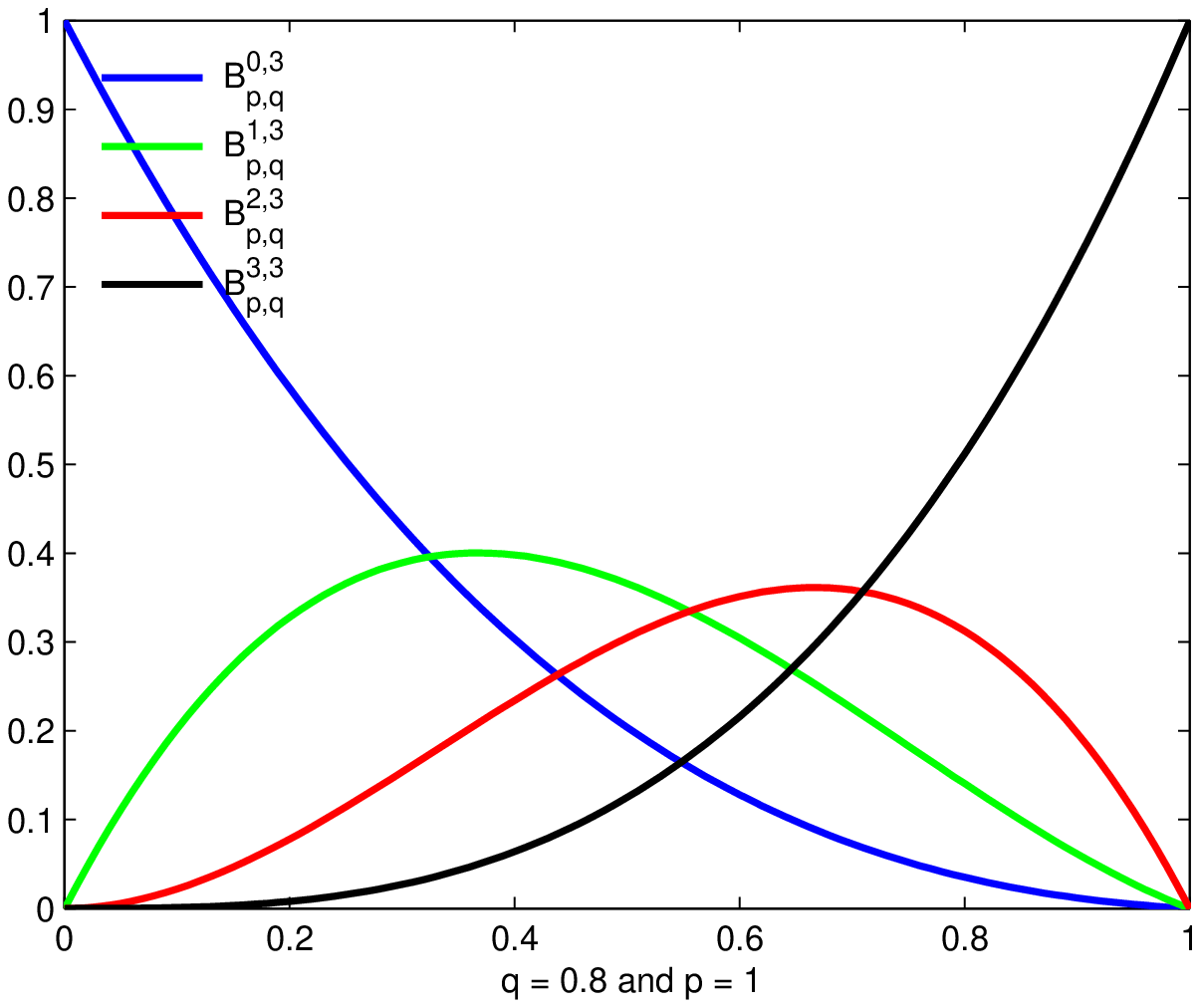}
\end{center}
\caption{`Cubic Bezier blending functions'}\label{f1}
\end{figure*}

\begin{figure*}[htb!]
\begin{center}
\includegraphics[height=6cm, width=8cm]{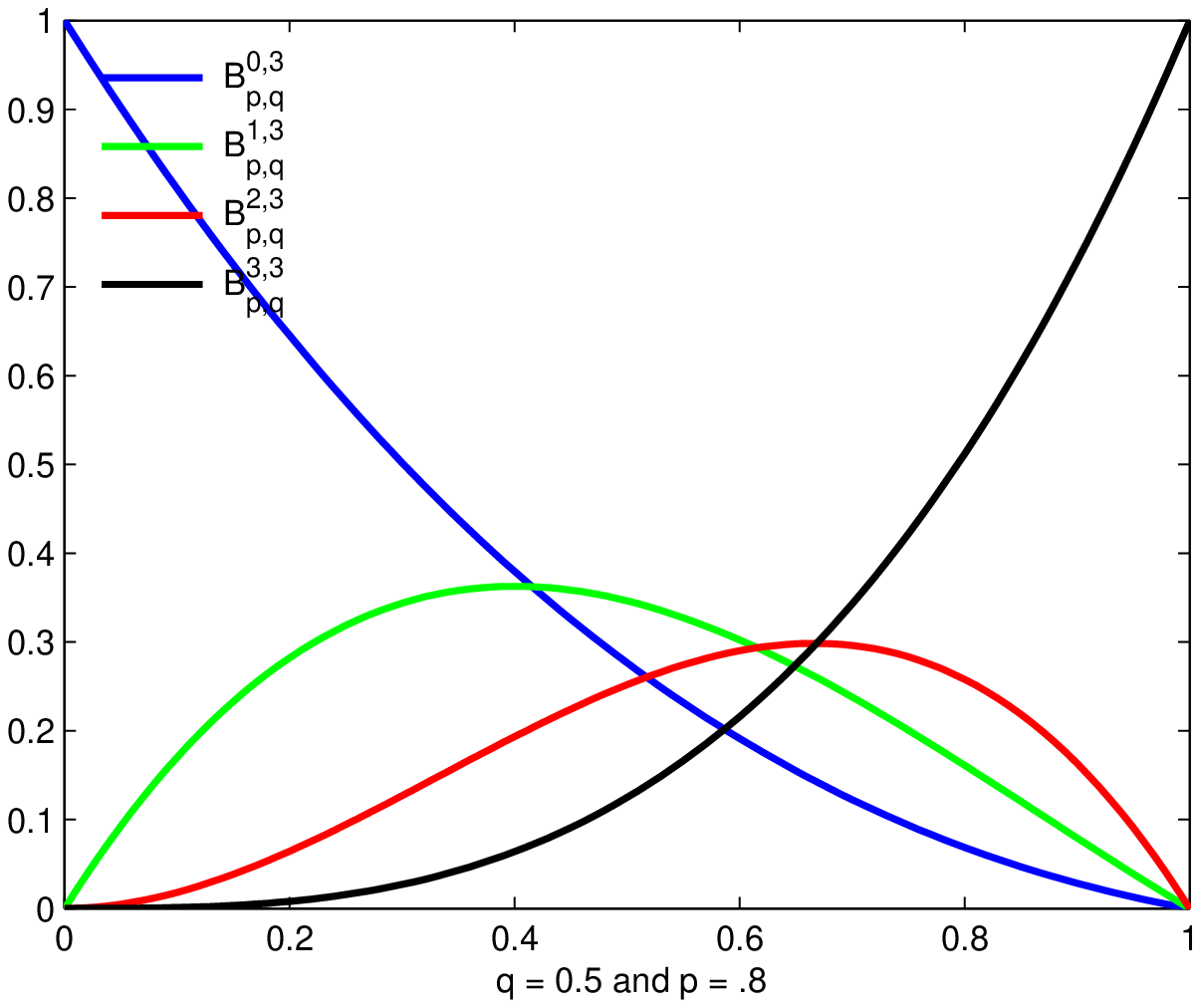}
\end{center}
\caption{`Cubic Bezier blending functions'}\label{f2}
\end{figure*}
\newpage

Fig. $\ref{f2}$ shows the $(p,q)$-analogues of the Bernstein basis functions of degree $3$ with $q = 0.5, p=.8 $. Here we can observe that sum of blending  fuctions is always unity and also end point interpolation property holds, when we put $p=1,$ it turns out to be $q$-Bernstein basis which is shown in Fig. $\ref{f1}.$\\

Apart from the basic properties above, the $(p,q)$-analogue of the Bernstein functions also satisfy some
recurrence relations, as for the classical Bernstein basis.\\

\section {Degree elevation and reduction for $(p,q)$-Bernstein functions}
Technique of degree elevation has been used to increase the flexibility of a given curve.
A degree elevation algorithm calculates a new set of control points by choosing a convex combination
of the old set of control points which retains the old end points. For this purpose, the identities (\ref{e6}),(\ref{e7}) and Theorem (\ref{t2}) are useful.\\

 \begin{thm}\label{t1}

  Each $(p,q)$-Bernstein functions of degree n is a linear combination of two $(p,q)$-Bernstein functions of degree $n-1:$
\begin{equation}\label{e5}
B^{k,n}_{p,q}(t)= q^{n-k}p^{k-1}~t~B^{k-1,n-1}_{p,q}(t)+(p^{n-1}-p^{k}q^{n-k-1} t)~B^{k,n-1}_{p,q}(t)
\end{equation}

\begin{equation}\label{e5a}
B^{k,n}_{p,q}(t)= p^{n-1} ~t~B^{k-1,n-1}_{p,q}(t)+(q^k p^{n-k-1}-q^{n-1} t)~B^{k,n-1}_{p,q}(t)
\end{equation}

%
%

\end{thm}

\textbf{Proof:}
 On using  Pascal's type relation  based on $(p,q)$-integers i.e $(\ref{e2}),$ we get
\begin{equation*}
B^{k,n}_{p,q}(t)=\frac{1}{p^{\frac{n(n-1)}{2}}}\bigg(q^{n-k}\left[
\begin{array}{c}
n-1 \\
k-1%
\end{array}%
\right] _{p,q}+p^k\left[
\begin{array}{c}
n-1 \\
k%
\end{array}%
\right] _{p,q}\bigg)~p^{\frac{k(k-1)}{2}}~~t^{k}(1-t)^{n-k}_{p,q}
\end{equation*}\\

\begin{align*}
&=\frac{1}{p^{\frac{n(n-1)}{2}}}q^{n-k}\left[
\begin{array}{c}
n-1 \\
k-1%
\end{array}%
\right] _{p,q} p^{\frac{k(k-1)}{2}}~~t^{k}(1-t)^{n-k}_{p,q}+\frac{1}{p^{\frac{n(n-1)}{2}}} p^k \left[
\begin{array}{c}
n-1 \\
k%
\end{array}%
\right] _{p,q}~p^{\frac{k(k-1)}{2}}t^{k}(1-t)^{n-k}_{p,q}\\
&=\frac{1}{p^{\frac{n(n-1)}{2}}}q^{n-k}p^{k-1}~~t\left[
\begin{array}{c}
n-1 \\
k-1%
\end{array}%
\right] _{p,q}p^{\frac{(k-1)(k-2)}{2}}t^{k-1}(1-t)^{n-k}_{p,q}\\
&~~+\frac{1}{p^{\frac{n(n-1)}{2}}}~~p^k (p^{n-k-1}-q^{n-k-1}t)\left[
\begin{array}{c}
n-1 \\
k%
\end{array}%
\right] _{p,q}~p^{\frac{k(k-1)}{2}}t^{k}(1-t)^{n-k-1}_{p,q}\\
&= q^{n-k}p^{k-1}~t~B^{k-1,n-1}_{p,q}(t)+(p^{n-1}-p^{k}q^{n-k-1} t)~B^{k,n-1}_{p,q}(t)
\end{align*}
Thus
\begin{equation*}
B^{k,n}_{p,q}(t)= q^{n-k}p^{k-1}~t~B^{k-1,n-1}_{p,q}(t)+(p^{n-1}-p^{k}q^{n-k-1} t)~B^{k,n-1}_{p,q}(t)
\end{equation*}

Similarly, if we use $(\ref{e3})$, we have

\begin{equation*}
B^{k,n}_{p,q}(t)= p^{n-1} ~t~B^{k-1,n-1}_{p,q}(t)+(q^k p^{n-k-1}-q^{n-1} t)~B^{k,n-1}_{p,q}(t)
\end{equation*}\\

\textbf{Degree elevation}

\begin{equation}\label{e6}
 q^{n-k}p^k~t~B^{k,n}_{p,q}(t)=\bigg(1-\frac{p^{k+1}~{[n-k]}_{p,q}}{{[n+1]}_{p,q}}\bigg)~B^{k+1,n+1}_{p,q}(t)
\end{equation}\\

\begin{equation}\label{e7}
(p^n- p^kq^{n-k}~t)~B^{k,n}_{p,q}(t)=\bigg(\frac{p^{k}~{[n+1-k]}_{p,q}}{{[n+1]}_{p,q}}\bigg)~B^{k,n+1}_{p,q}(t)
\end{equation}\\

\textbf{Proof:}
\begin{align*}
q^{n-k}p^k~t~B^{k,n}_{p,q}(t)&= \frac{1}{p^{\frac{n(n-1)}{2}}} q^{n-k}p^k~t~\bigg(\left[
\begin{array}{c}
n \\
k%
\end{array}%
\right] _{p,q}p^{\frac{k(k-1)}{2}}t^{k}~(1-t)^{n-k}_{p,q}\bigg)\\
&=q^{n-k} \frac{{[k+1]}_{p,q}}{{[n+1]}_{p,q}}\frac{1}{p^{\frac{n(n-1)}{2}}}~\bigg(\left[
\begin{array}{c}
n+1 \\
k+1%
\end{array}%
\right] _{p,q} p^{\frac{(k+1)(k)}{2}} t^{k+1}(1-t)^{n-k}_{p,q}\bigg)\\
&=q^{n-k} \frac{{[k+1]}_{p,q}}{{[n+1]}_{p,q}}~B^{k+1,n+1}_{p,q}(t)
\end{align*}

By some simple calculation, we have
\begin{equation*}
q^{n-k} \frac{{[k+1]}_{p,q}}{{[n+1]}_{p,q}}=~1-\frac{p^{k+1}~{[n-k]}_{p,q}}{{[n+1]}_{p,q}},
\end{equation*}
using this result, we get

\begin{equation*}
 q^{n-k} p^k~t~B^{k,n}_{p,q}(t)=\bigg(1-\frac{p^{k+1}~{[n-k]}_{p,q}}{{[n+1]}_{p,q}}\bigg)~B^{k+1,n+1}_{p,q}(t),
\end{equation*}
similarly on considering,
\begin{align*}
(p^n- p^k q^{n-k}~t)~B^{k,n}_{p,q}(t)&= (p^n- p^k q^{n-k}~t)~\bigg(\frac{1}{p^{\frac{n(n-1)}{2}}}\left[
\begin{array}{c}
n \\
k%
\end{array}%
\right] _{p,q} p^{\frac{k(k-1)}{2}} t^{k}~(1-t)^{n-k}_{p,q}\bigg)\\
&=\frac{(p^n- p^k q^{n-k}~t)}{(p^{n-k}- q^{n-k}~t)}~\frac{1}{p^{\frac{n(n-1)}{2}}} \frac{{[n+1-k]}_{p,q}}{{[n+1]}_{p,q}}  \bigg(\left[
\begin{array}{c}
n+1 \\
k%
\end{array}%
\right] _{p,q} p^{\frac{k(k-1)}{2}} t^{k}~(1-t)^{n+1-k}_{p,q}\bigg)
\end{align*}

finally we get
\begin{equation*}
(p^n- p^kq^{n-k}~t)~B^{k,n}_{p,q}(t)= \bigg(\frac{p^{k}~{[n+1-k]}_{p,q}}{{[n+1]}_{p,q}}\bigg)~B^{k,n+1}_{p,q}(t)
\end{equation*}\\

 \begin{thm}\label{t2}
Each $(p,q)$-Bernstein function of degree $n$ is a linear combination of two $(p,q)$-Bernstein functions of degree $n+1.$
\begin{equation}\label{e8}
B^{k,n}_{p,q}(t)= \bigg(\frac{p^{k-n}~{[n+1-k]}_{p,q}}{{[n+1]}_{p,q}}\bigg)~B^{k,n+1}_{p,q}(t)+p^{-n}\bigg(1-\frac{p^{k+1}~{[n-k]}_{p,q}}{{[n+1]}_{p,q}}\bigg)~B^{k+1,n+1}_{p,q}(t)
\end{equation}
\end{thm}

\textbf{Proof} From  equation $\ref{e6},\ref{e7}$ we can easily get

\begin{equation*}
B^{k,n}_{p,q}(t)= \bigg(\frac{p^{k-n}~{[n+1-k]}_{p,q}}{{[n+1]}_{p,q}}\bigg)~B^{k,n+1}_{p,q}(t)+ p^{-n}\bigg(1-\frac{p^{k+1}~{[n-k]}_{p,q}}{{[n+1]}_{p,q}}\bigg)~B^{k+1,n+1}_{p,q}(t).
\end{equation*}\\

For  $n=3,$  the  blending functions are given by:

 $$ B_{p,q}^{0,3} = \frac{1}{p^3} \left[
\begin{array}{c}
3 \\
0
\end{array}%
\right] _{p,q} (1-t) (p-qt) (p^2-q^2t)$$
  $$B_{p,q}^{1,3} = \frac{1}{p^3} \left[
\begin{array}{c}
3 \\
1
\end{array}%
\right] _{p,q}  t (1-t) (p-qt)$$
   $$B_{p,q}^{2,3} = \frac{1}{p^3} \left[
\begin{array}{c}
3 \\
2
\end{array}%
\right] _{p,q} p t^2 (1-t)$$
    $$ B_{p,q}^{3,3} =  \left[
\begin{array}{c}
3 \\
3\end{array}%
\right] _{p,q}  t^3$$
\newline
We observed that the both sided end point interpolation property and partition of unity property always holds in case of $(p,q)$-Bernstein functions.\\

The de Casteljau algorithm describes how to subdivide a B$\acute{e}$zier curve, when a B$\acute{e}$zier curve is repeatedly subdivided, the collection of control polygons converge to the curve. Thus, the way of computing a B$\acute{e}$zier curve is to simply subdivide it an appropriate number of times and compute the control polygons.

\section{ $(p,q)$-Bernstein B$\acute{e}$zier curves:}

Let us define the $(p,q)$-B$\acute{e}$zier curves of degree n using the $(p,q)$-analogues of the Bernstein functions as follows:

\begin{equation}\label{e12}
{\bf{ P}}(t; p,q) = \sum\limits_{i=0}^{n} {\bf{P_i}}~B^{i,n}_{p,q}(t)
\end{equation}

where $P_i \in R^3$  $(i = 0, 1, . . . , n)$ and $ p>q > 0.$ $P_i$ are control points. Joining up adjacent points $P_i,$  $i = 0, 1, 2, . . . , n $ to obtain a
polygon which is called the control polygon of $(p,q)$-Bézier curves.\\

\subsection{ Some basic properties of $(p,q)$-B$\acute{e}$zier curves.}

 \begin{thm} From the definition, we can derive some basic properties of $(p,q)$-B$\acute{e}$zier curves:\\

\noindent 1.  $(p,q)$-B$\acute{e}$zier curves have geometric and affine invariance.\\
2.  $(p,q)$-B$\acute{e}$zier curves lie inside the convex hull of its control polygon.\\
3. The end-point interpolation property: ${\bf{P}}(0; p, q) = {\bf{P_0,}} ~ {\bf{P}}(1; p, q) = \bf{P_n.}$\\
3. Reducibility: when $p = 1,$ formula \ref{e12} gives the $q$-B$\acute{e}$zier curves.
\end{thm}

\textbf{Proof.} These properties of  $(p,q)$-B$\acute{e}$zier curves can be easily deduced from corresponding properties of the
$(p,q)$-analogue of the Bernstein functions.

%
%
%
%

 \subsection{Degree elevation for $(p,q)$-B$\acute{e}$zier curves}

 $(p,q)$-B$\acute{e}$zier curves have a degree elevation algorithm that is similar to that possessed by the classical B$\acute{e}$zier curves.
Using the technique of degree elevation, we can increase the flexibility of a given curve.

 $$ {\bf{ P}}(t; p,q) = \sum\limits_{k=0}^{n} {\bf{P_k}}~ B^{k,n}_{p,q}(t)$$

 $$ {\bf{ P}}(t; p,q) = \sum\limits_{k=0}^{n+1} {\bf{P_k^\ast}}~ B^{i+1,n}_{p,q}(t),$$ where

\begin{equation}\label{e15}
{\bf{ P^\ast}}=p^{-n}\bigg(1-\frac{p^{k}~{[n+1-k]}_{p,q}}{{[n+1]}_{p,q}}\bigg)~{\bf{ P_{k-1}}}+ p^{k-n} \bigg(\frac{~{[n+1-k]}_{p,q}}{{[n+1]}_{p,q}}\bigg)~{\bf{ P_k}}
\end{equation}

The statement above can be derived using the identities $(\ref{e6}) \text{and} (\ref{e7}).$
Consider

$$ p^n{\bf{ P}}(t; p,q)= (p^n-p^k q^{n-k}t)~~ {\bf{ P}}(t; p,q) + p^k q^{n-k}t ~~{\bf{ P}}(t; p,q).$$
We obtain

\begin{equation*}
p^n{\bf{P}}(t; p,q)=\sum\limits_{k=0}^{n} \bigg(p^k ~~\frac{~{[n+1-k]}_{p,q}}{{[n+1]}_{p,q}}\bigg) {\bf{P^0_k}} B^{k,n+1}_{p,q}(t) + \sum\limits_{k=0}^{n}\bigg(1-\frac{p^{k+1}~{[n-k]}_{p,q}}{{[n+1]}_{p,q}}\bigg){\bf{P^0_k}} B^{k+1,n+1}_{p,q}(t)
\end{equation*}

Now by shifting the limits, we have

\begin{equation*}
p^n{\bf{P}}(t; p,q)=\sum\limits_{k=0}^{n+1} \bigg(p^k\frac{~{[n+1-k]}_{p,q}}{{[n+1]}_{p,q}}\bigg) {\bf{P^0_k}} B^{k,n+1}_{p,q}(t) + \sum\limits_{k=0}^{n+1}\bigg(1-\frac{p^{k}~{[n+1-k]}_{p,q}}{{[n+1]}_{p,q}}\bigg){\bf{P^0_{k-1}}} B^{k,n+1}_{p,q}(t)
\end{equation*}

where ${\bf{P^0_{-1}}} $is defined as the zero vector.
Comparing coefficients on both side, we have

\begin{equation*}
p^n{\bf{P^1_k}}= \bigg(p^k\frac{~{[n+1-k]}_{p,q}}{{[n+1]}_{p,q}}\bigg) {\bf{P^0_k}}  + \bigg(1-\frac{p^{k}~{[n+1-k]}_{p,q}}{{[n+1]}_{p,q}}\bigg){\bf{P^0_{k-1}}}
\end{equation*}

or

\begin{equation*}
{\bf{P^1_k}}= \bigg( p^{k-n} \frac{~{[n+1-k]}_{p,q}}{{[n+1]}_{p,q}}\bigg) {\bf{P^0_k}}  + p^{-n}\bigg (1-\frac{p^{k}~{[n+1-k]}_{p,q}}{{[n+1]}_{p,q}}\bigg){\bf{P^0_{k-1}}}.
\end{equation*}
In general

\begin{equation*}
{\bf{P^r_k}}= \bigg( p^{k-n} \frac{~{[n+r-k]}_{p,q}}{{[n+r]}_{p,q}}\bigg) {\bf{P^{r-1}_k}}  + p^{-n}\bigg (1-\frac{p^{k}~{[n+1-k]}_{p,q}}{{[n+1]}_{p,q}}\bigg){\bf{P^{r-1}_{k-1}}}
\end{equation*}
where $r=1,.....n~~ \text{and}~~ i=0,1,2,...,n+r$

 When $ p = 1,$ formula \ref{e15} reduce to the degree evaluation formula
of the $q$-B$\acute{e}$zier curves. If we let $ P = (P_0, P_1, . . . , P_n)^{T}$  denote the vector of control points of the initial $(p,q)$-B$\acute{e}$zier
curve of degree $n,$ and $ {\bf{P^{(1)}}}=(P_0^\ast, P_1^\ast, . . . , P_{n+1}^\ast)$
 the vector of control points of the degree elevated $(p,q)$-B$\acute{e}$zier curve of
degree $n + 1,$ then we can represent the degree elevation procedure as:

$${\bf{P^{(1)}}}=T_{n+1}{\bf{P}},$$

where

 $$T_{n+1}=\dfrac{1}{[n+1]_{p,q}}\begin{bmatrix}
\;\frac{[n+1]_{p,q}} {p^{n}}\; & \;0\; & \;\ldots \; & \;0\; & \;0\; \\
\frac{([n+1]_{p,q}-p[1]_{p,q})}{ p^{n}}& \frac{[1] _{p,q}}{ p^{n-2}} & \ldots & 0 & 0 \\
\vdots & \vdots & \ddots & \vdots & \vdots \\
0 & \ldots &\frac{([n+1] _{p,q} -p^{2}[n-1] _{p,q})}  {p^{-n}} &\frac{ [n-1] _{p,q}}{ p^{n-4}} & 0 \\
\;0\; & \;0\; & \;\ldots & \;0\; & \;\frac{[n+1] _{p,q}}{p^{n}} \;
\end{bmatrix}_{(n+2)\times(n+1)}$$

For any $ l \in \mathbb{N},$ the vector of control points of the degree elevated  $(p,q)$-Bézier curves of degree $ n + l$ is:
${\bf{P^{(l)}}} = T_{n+l}~ T_{n+2}........ T_{n+1} {\bf{P}}.$
As $l \longrightarrow \infty,$ the control polygon $\bf{P^{(l)}}$ converges to a $(p,q)$-B$\acute{e}$zier curve.

\subsection{de Casteljau algorithm:}

 $(p,q)$-B$\acute{e}$zier curves of degree $n$ can be written as two kinds of linear combination of two  $(p,q)$-B$\acute{e}$zier curves of degree $n-1,$ and we can get the two selectable algorithms to evaluate  $(p,q)$-B$\acute{e}$zier curves. The algorithms can be expressed as:

\textbf{ Algorithm 1.}\\

\begin{equation}\label{e14}
\left\{
 \begin{array}{ll}
 {\bf{P^{0}_{i}}}(t;p,q)\equiv {\bf{P^{0}_{i}}}\equiv {\bf{P_{i}}}~~~i=0,1,2......,n~~~\mbox{ } &  \\
 &  \\
{\bf{P^{r}_{i}}}(t;p,q)= p^{r-1}t~{\bf{P^{r-1}_{i+1}}}(t;p,q)+ (q^k p^{r-i-1}-q^{r-1}t)~{\bf{P^{r-1}_{i}}}(t;p,q)~~~\mbox{  } &\\
 r=1,...,n,~~~i=0,1,2......,n-r.,~~~\mbox{  }
 \end{array}%
 \right.
 \end{equation}
or

\begin{equation}\label{e15}
\left\{
 \begin{array}{ll}
 {\bf{P^{0}_{i}}}(t;p,q)\equiv {\bf{P^{0}_{i}}}\equiv {\bf{P_{i}}}~~~i=0,1,2......,n~~~\mbox{ } &  \\
 &  \\
{\bf{P^{r}_{i}}}(t;p,q)= p^{i} q^{r-k-1}t~{\bf{P^{r-1}_{i+1}}}(t;p,q)+ ( p^{r-1}- p^k q^{r-i-1}t)~{\bf{P^{r-1}_{i}}}(t;p,q)~~~~\mbox{  } &\\
 r=1,...,n,~~~i=0,1,2......,n-r.,~~~\mbox{  }
 \end{array}%
 \right.
 \end{equation}

Then

 \begin{equation}\label{}
  {\bf{ P}}(t; p,q) = \sum\limits_{i=0}^{n-1} {\bf{P_i^1}}(t; p,q)=...=\sum\limits {\bf{P_i^r}}(t; p,q)~B^{i,n-r}_{p,q}(t)=...= {\bf{P_0^n}}~(t; p,q)
\end{equation}

It is clear that the results can be obtained from Theorem (\ref{khalidthm}).  When $p = 1,$ formula (\ref{e14}) and (\ref{e15}) recover the de Casteljau algorithms of classical $q$-B$\acute{e}$zier curves. Let $P^0 = (P_0, P_1, . . . , P_n)^T$ , $P^r = (P_0^r,P_1^r,....,P_{n-r}^r)^{T},$
 then de Casteljau algorithm can be expressed as:\\

\textbf{ Algorithm 2.}

\begin{equation}\label{e16}
 {\bf{ P^r}}(t; p,q)=M_r(t; p,q)....M_2(t; p,q)M_1(t; p,q){\bf{ P^0}}
\end{equation}
where $M_r(t; p,q)$ is a $(n - r + 1) \times (n - r + 2) $ matrix and

$$ M_r(t; p,q)=\begin{bmatrix}
\;  (p^{r-1}-q^{r-1}t)\; & \; p^{r-1} t \; & \;\ldots \; & \;0\; & \;0\; \\
0 &  (qp^{r-2}-q^{r-1}t)    & p^{r-1} t & 0 & 0 \\
\vdots & \vdots & \ddots & \vdots & \vdots \\
0 & \ldots & (q^{n-r-1}p^{2r-n-2}-q^{r-1}t)      & p^{r-1} t  & 0 \\
0 & 0 & \ldots & (q^{n-r}p^{2r-n-1}-q^{r-1}t) & p^{r-1} t
\end{bmatrix}$$
or
$$ M_r(t; p,q)=\begin{bmatrix}
\; (p^{r-1}-q^{r-1}t)  \; & \; q^{r-1}t \; & \;\ldots \; & \;0\; & \;0\; \\
0 & (p^{r-1}-pq^{r-2}t)  & pq^{r-2}t  & 0 & 0 \\
\vdots & \vdots & \ddots & \vdots & \vdots \\
0 & \ldots & ~~ &~~ & 0 \\
0 & 0 & \ldots & (p^{r-1}-p^{n-r}q^{-n-1}t) & p^{n-r}q^{-n-1}t
\end{bmatrix}$$

\section{Tensor product $(p,q)$-Bernstein B$\acute{e}$zier surfaces on $[0, 1] \times  [0, 1]$}

We define a two-parameter family ${{\bf{P}}}(u,v)$ of tensor product surfaces of degree $m \times n$ as follow:

\begin{equation}\label{e18}
{{\bf{P}}}(u,v) = \sum\limits_{i=0}^{m}\sum\limits_{j=0}^{n} {{\bf{P}}_{i,j}}~ B^{i,m}_{p_1,q_1}(u)~~ B^{j,n}_{p_2,q_2}(v),~~(u,v) \in [0, 1] \times  [0, 1],
\end{equation}

where ${{\bf{P}}_{i,j}} \in \mathbb{R}^3 ~~(i = 0, 1, . . . ,m, j = 0, 1, . . . , n)$ and two real numbers $p_1> q_1>0,~ p_2> q_2 > 0,$ $b^{i,m}_{p_1,q_1}(u),~~ b^{j,n}_{p_2,q_2}(v)$ are $(p,q)$-analogue of Bernstein functions respectively with the parameter $ p_1,q_1$ and $ p_2, q_2.$ We call the parameter surface tensor product
 $(p,q)$-B$\acute{e}$zier surface with degree $m \times n.$ We refer to the ${{\bf{P}}_{i,j}}$ as the control points. By joining up adjacent points in the same
row or column to obtain a net which is called the control net of tensor product $(p,q)$-B$\acute{e}$zier surface.\\

\subsection{Properties}

1. \textbf{Geometric invariance and affine invariance property:} Since
\begin{equation}\label{}
 \sum\limits_{i=0}^{m}\sum\limits_{j=0}^{n}  B^{i,m}_{p_1,q_1}(u)~~ B^{j,n}_{p_2,q_2}(v)=1,
\end{equation}

 ${{\bf{P}}}(u,v)$ is an affine combination
of its control points.\\

2. \textbf{Convex hull property:} ${{\bf{P}}}(u,v)$ is a convex combination of ${{\bf{P}}_{i,j}}$ and lies in the convex hull of its control net.\\

3. \textbf{Isoparametric curves property:} The isoparametric curves $v = v^\ast$ and $u = u^\ast$ of a tensor product $(p,q)$-B$\acute{e}$zier surface
are respectively the  $(p,q)$-B$\acute{e}$zier curves of degree $m$ and degree $n,$ namely,

\begin{equation*}
{{\bf{P}}}(u,v^\ast) = \sum\limits_{i=0}^{m}\bigg(\sum\limits_{j=0}^{n} {{\bf{P}}_{i,j}}~B^{j,n}_{p_2,q_2}(v^\ast)\bigg)~ B^{i,m}_{p_1,q_1}(u)~~ ,~~u  \in [0, 1] ;
\end{equation*}

\begin{equation*}
{{\bf{P}}}(u^\ast,v) = \sum\limits_{j=0}^{n}\bigg(\sum\limits_{i=0}^{m} {{\bf{P}}_{i,j}}~B^{j,n}_{p_1,q_1}(u^\ast)\bigg)~ B^{i,m}_{p_2,q_2}(v)~~ ,~~v  \in [0, 1]
\end{equation*}

The boundary curves of ${{\bf{P}}}(u,v)$ are evaluated by ${{\bf{P}}}(u,0)$, ${{\bf{P}}}(u,1)$, ${{\bf{P}}}(0,v)$ and ${{\bf{P}}}(1,v)$.\\

4. \textbf{Corner point interpolation property: }The corner control net coincide with the four corners of the surface. Namely, ${{\bf{P}}}(0,0)={{\bf{P}}}_{0,0},$
${{\bf{P}}}(0,1) ={{\bf{P}}}_{0,n},$
 ${{\bf{P}}}(1,0) ={{\bf{P}}}_{m,0},$
 ${{\bf{P}}}(1,1) ={{\bf{P}}}_{m,n},$\\

5. \textbf{Reducibility:} When $p_1 = p_2 = 1,$ formula (\ref{e18}) reduces to a tensor product $q$-B$\acute{e}$zier patch.

\subsection{Degree elevation and de Casteljau algorithm}
Let ${{\bf{P}}}(u,v)$ be a tensor product $(p,q)$-B$\acute{e}$zier surface of degree $m \times n.$ As an example, let us take obtaining the same
surface as a surface of degree $(m + 1) \times (n + 1).$ Hence we need to find new control points ${{\bf{P}}}_{i,j}^\ast$ such that

\begin{equation}\label{}
{{\bf{P}}}(u,v) = \sum\limits_{i=0}^{m}\sum\limits_{j=0}^{n} {{\bf{P}}_{i,j}}~ B^{i,m}_{p_1,q_1}(u)~~ B^{j,n}_{p_2,q_2}(v)= \sum\limits_{i=0}^{m+1}\sum\limits_{j=0}^{n+1} {\bf{P^\ast}_{i,j}}~ B^{i,m+1}_{p_1,q_1}(u)~~ B^{j,n+1}_{p_2,q_2}(v)
\end{equation}
Let
$\alpha_i=p_1^{-m}\bigg(1-\frac{p_1^{i}~{[m+1-i]}_{p_1,q_1}}{{[m+1]}_{p_1,q_1}}\bigg),~~$ $\beta_j=p_2^{-n}\bigg(1-\frac{p_2^{j-1}~{[n+1-j]}_{p_2,q_2}}{{[n+1]}_{p_2,q_2}}\bigg).$

Then
\begin{equation}\label{}
  {{\bf{P}}}_{i,j}^\ast=\alpha_i~\beta_j~{{\bf{P}}}_{i-1,j-1}+\alpha_i~(1-\beta_j)~{{\bf{P}}}_{i-1,j}+(1-\alpha_i)~(1-\beta_j)~{{\bf{P}}}_{i,j}
\end{equation}

which can be written in matrix form as
$$
  \begin{bmatrix}
   p_1^{-m} \bigg(1-\frac{p_1^{i}~{[m+1-i]}_{p_1,q_1}}{{[m+1]}_{p_1,q_1}}\bigg) & \frac{p_1^{i-m}~{[m+1-i]}_{p_1,q_1}}{{[m+1]}_{p_1,q_1}}\\
  \end{bmatrix}
    \begin{bmatrix}
      {{\bf{P}}}_{i-1,j-1} & {{\bf{P}}}_{i-1,j} \\
     {{\bf{P}}}_{i,j-1} & {{\bf{P}}}_{i,j} \\
    \end{bmatrix}
                   \begin{bmatrix}
                     p_2^{-n} \bigg( 1-\frac{p_2^{j-1}~{[n+1-j]}_{p_2,q_2}}{{[n+1]}_{p_2,q_2}}\bigg) \\
                     \frac{p_2^{j-n}~{[n+1-j]}_{p_2,q_2}}{{[n+1]}_{p_2,q_2}}\\
                   \end{bmatrix}
    $$

The de Casteljau algorithms are also easily extended to evaluate points on a $(p,q)$-B$\acute{e}$zier surface. Given the control net
${{\bf{P}}}_{i,j} \in \mathbb{R}^3, i = 0, 1, . . . ,m,~~ j = 0, 1, . . . , n.$

\begin{equation}\label{e21}
\left\{
 \begin{array}{ll}
 {\bf{P^{0,0}_{i,j}}}(u,v)\equiv {\bf{P^{0,0}_{i,j}}}\equiv {\bf{P_{i,j}}}~~~i=0,1,2......,m;~~j=0,1,2...n.\mbox{ } &  \\
 &  \\
{\bf{P^{r,r}_{i,j}}}(u,v)= \begin{bmatrix}
    (q_1^i p_1^{r-i-1} -q_1^{r-1}u)~~ & p_1^{r-1}u\\
  \end{bmatrix}
    \begin{bmatrix}
      {{\bf{P}}}_{i,j}^{r-1,r-1} & {{\bf{P}}}_{i,j+1}^{r-1,r-1} \\
     {{\bf{P}}}_{i+1,j}^{r-1,r-1} & {{\bf{P}}}_{i+1,j+1}^{r-1,r-1} \\
    \end{bmatrix}
                   \begin{bmatrix}
                      (q_2^j p_2^{r-j-1} -q_2^{r-1}v) \\
                     p_2^{r-1}v\\
                   \end{bmatrix}      ~~~\mbox{  } &\\
 r=1,...,k, ~k=\text{min}(m,n)~~~i=0,1,2......,m-r;~~ j=0,1,....n-r~~~\mbox{  }
 \end{array}%
 \right.
 \end{equation}
 or

\begin{equation}\label{e22}
\left\{
 \begin{array}{ll}
 {\bf{P^{0,0}_{i,j}}}(u,v)\equiv {\bf{P^{0,0}_{i,j}}}\equiv {\bf{P_{i,j}}}~~~i=0,1,2......,m;~~j=0,1,2...n.\mbox{ } &  \\
 &  \\
{\bf{P^{r,r}_{i,j}}}(u,v)= \begin{bmatrix}
    (p_1^{r-1}-p_1^iq_1^{r-i-1}u)~~ & p_1^{i} q_1^{r-i-1}u\\
  \end{bmatrix}
    \begin{bmatrix}
      {{\bf{P}}}_{i,j}^{r-1,r-1} & {{\bf{P}}}_{i,j+1}^{r-1,r-1} \\
     {{\bf{P}}}_{i+1,j}^{r-1,r-1} & {{\bf{P}}}_{i+1,j+1}^{r-1,r-1} \\
    \end{bmatrix}
                   \begin{bmatrix}
                      (p_2^{r-1}-p_2^j q_2^{r-j-1}v) \\
                     p_2^{j} q_2^{r-j-1}v\\
                   \end{bmatrix}      ~~~\mbox{  } &\\
 r=1,...,k, ~k=\text{min}(m,n)~~~i=0,1,2......,m-r;~~ j=0,1,....n-r~~~\mbox{  }
 \end{array}%
 \right.
 \end{equation}
 When $m = n,$ one can directly use the algorithms above to get a point on the surface. When $m \neq n,$ to get a point on the
surface after $k$ applications of formula (\ref{e21}) or (\ref{e22}), we perform formula (\ref{e16}) for the intermediate point $ {{\bf{P}}}_{i,j}^{k,k}.$\\

\textbf{Note:} We get $q$-B$\acute{e}$zier curves and surfaces for $(u, v) \in [0, 1] \times [0, 1] $  when we set the parameter $p_1=p_2=1$ as proved in \cite{cetin1,cetin2}.\\

\newpage

\section{Shape control of $(p,q)$-Bernstein curves}
We have constructed $(p,q)$-Bernstein functions which holds both the end point interpolation property as shown in figure $\ref{f1}$ and $\ref{f2}$. Parameter $p$ and $q$ has been used to control the shape of curves and surfaces: if $ 0 < q < p \leq 1,$ as $p$ and $q$ decreases, the curve moves close to the control polygon, as $p$ and $q$ increases, the curve moves far away from the control polygon; If $p>1$ and $q>1,$ the effects of $p$ and $q$ are opposite, as $p$ and $q$ decreases, the curve moves far away from the control polygon, as $p$ and $q$ increases, the curve moves close to the control polygon.\\
Figure $\ref{f3},\ref{f4}, \ref{f5}, \ref{f6} \text{and} \ref{f7}$ shows $(p,q)$-Bernstein function approximating the surface generated by control points.

\begin{figure*}[htb!]
\begin{center}
\includegraphics[height=6cm, width=8cm]{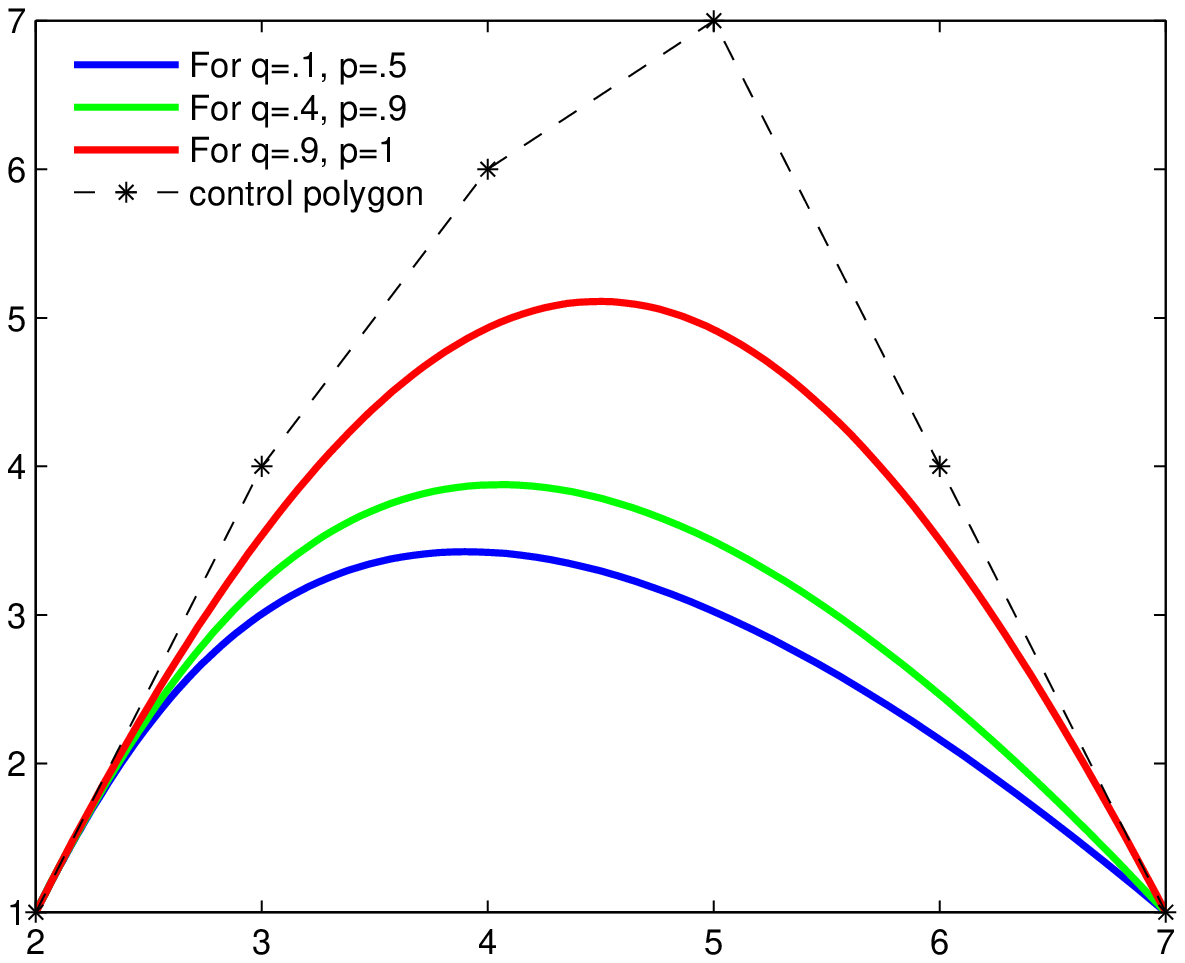}
\end{center}
\caption{`B$\acute{e}$zier curve}\label{f3}
\end{figure*}

\begin{figure*}[htb!]
\begin{center}
\includegraphics[height=6cm, width=8cm]{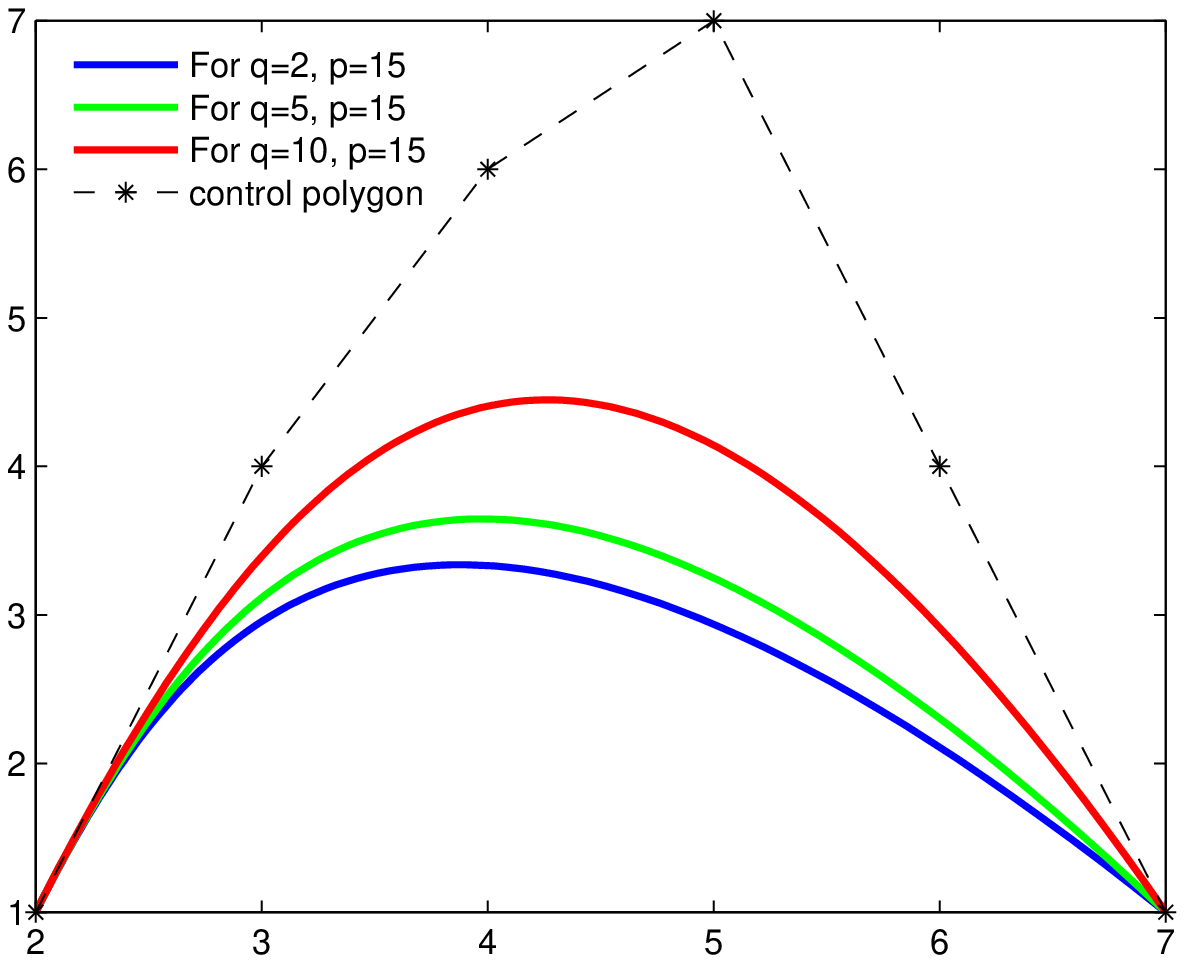}
\end{center}
\caption{`B$\acute{e}$zier curve}\label{f4}
\end{figure*}

\begin{figure*}[htb!]
\begin{center}
\includegraphics[height=6cm, width=8cm]{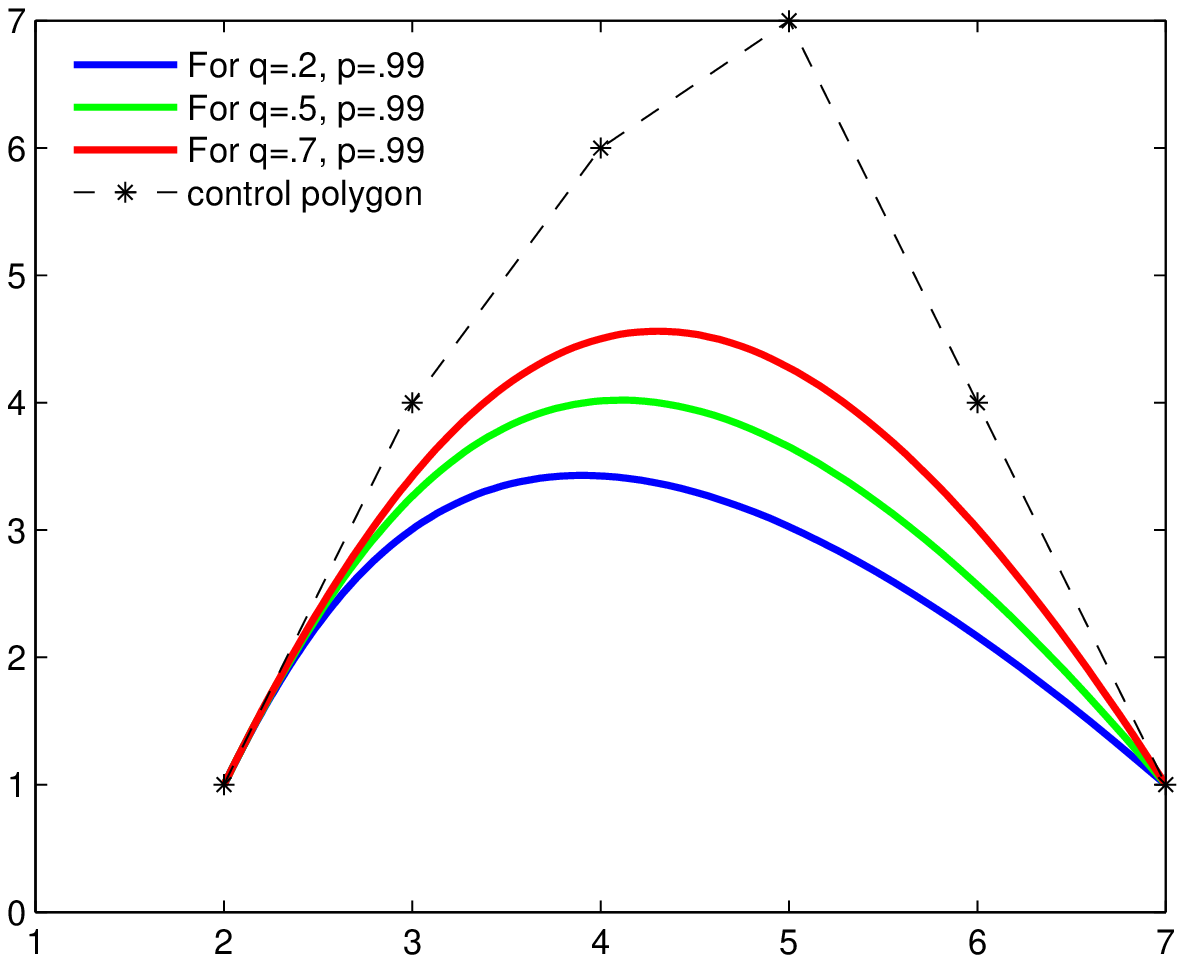}
\end{center}
\caption{`B$\acute{e}$zier curve'}\label{f5}
\end{figure*}

\begin{figure*}[htb!]
\begin{center}
\includegraphics[height=6cm, width=8cm]{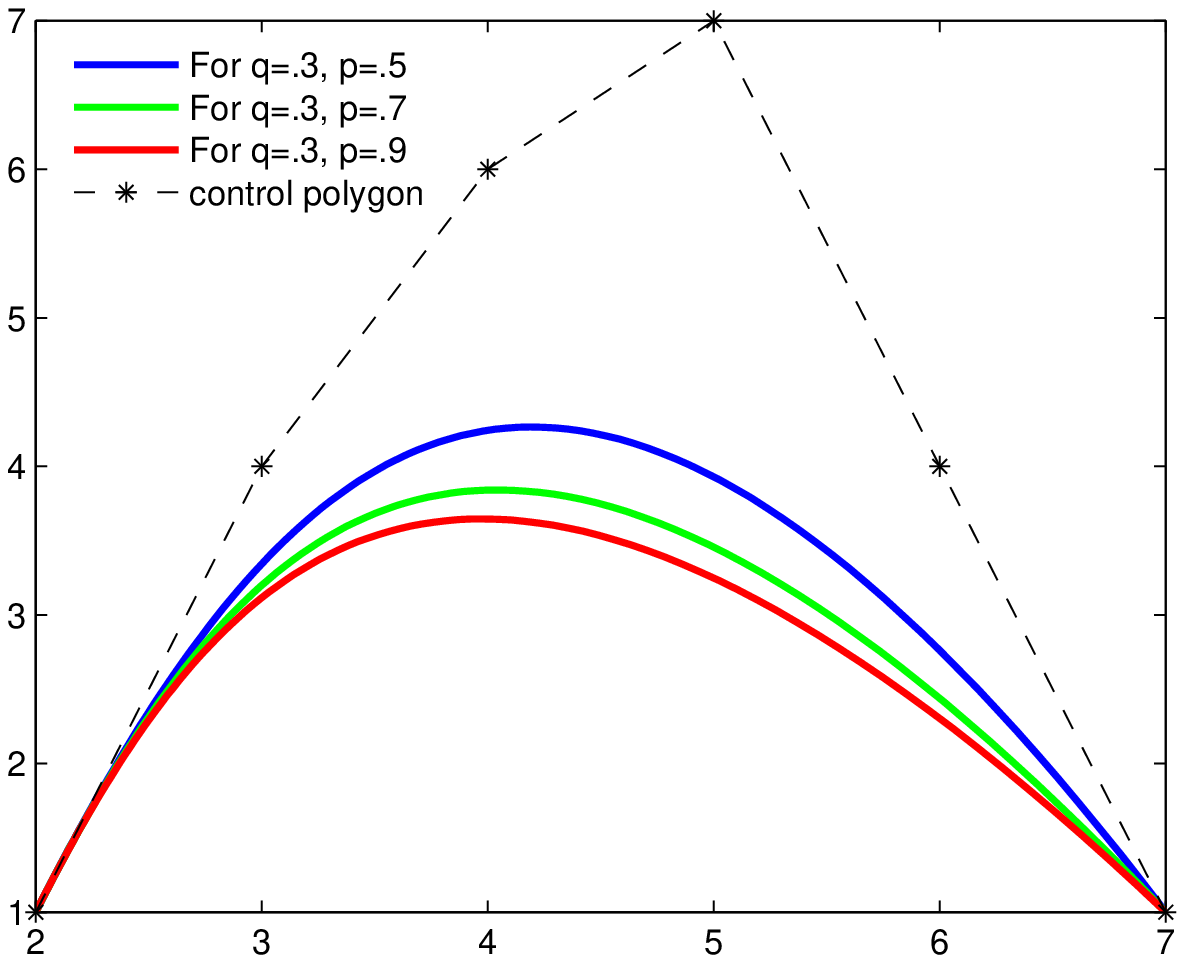}
\end{center}
\caption{`B$\acute{e}$zier curve'}\label{f6}
\end{figure*}

\begin{figure*}[htb!]
\begin{center}
\includegraphics[height=6cm, width=8cm]{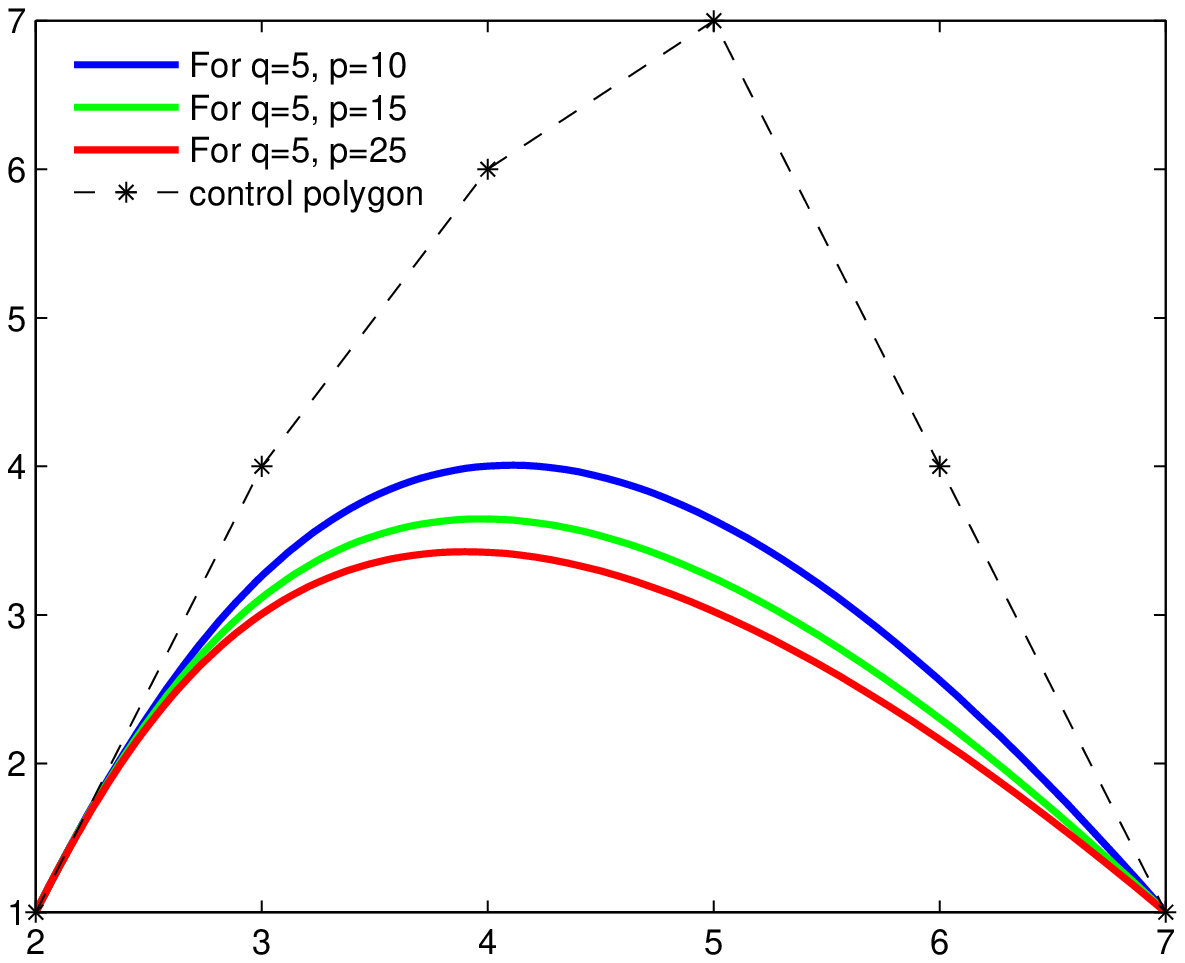}
\end{center}
\caption{`B$\acute{e}$zier curve}\label{f7}
\end{figure*}

%
}
%

\end{document}